# Nanoscale control of LaAlO$_3$/SrTiO$_3$ metal-insulator transition using ultra-low-voltage electron-beam lithography

Dengyu Yang[1, 4], Shan Hao[1, 4], Jun Chen[2], Qing Guo[1, 4], Muqing Yu[1, 4], Yang Hu[1, 4], Kitae Eom[3], Jung-Woo Lee[3], Chang-Beom Eom[3], Patrick Irvin[1, 4], Jeremy Levy[1, 4, a)]

**AFFILIATIONS**

1. Department of Physics and Astronomy, University of Pittsburgh, Pittsburgh, PA 15260, USA
2. Department of Electrical and Computer Engineering and Peterson Institute of Nanoscience and Engineering, University of Pittsburgh, Pittsburgh, PA 15261, USA
3. Department of Materials Science and Engineering, University of Wisconsin-Madison, Madison, WI 53706, USA
4. Pittsburgh Quantum Institute, Pittsburgh, PA 15260, USA
a) **Author to whom correspondence should be addressed:** jlevy@pitt.edu

## Abstract

We describe a method to control the insulator-metal transition at the LaAlO$_3$/SrTiO$_3$ interface using ultra-low-voltage electron beam lithography (ULV-EBL). Compared with previous reports that utilize conductive atomic-force-microscope lithography (c-AFM), this approach can provide comparable resolution (~10 nm) at write speeds (10 mm/s) that are up to 10,000x faster than c-AFM. The writing technique is non-destructive and the conductive state is reversible via prolonged exposure to air. Transport properties of representative devices are measured at milli-Kelvin temperatures, where superconducting behavior is observed. We also demonstrate the ability to create conducting devices on graphene/LaAlO$_3$/SrTiO$_3$ heterostructures. The underlying mechanism is believed to be closely related to the same mechanism regulating c-AFM-based methods.

## Introduction

The complex-oxide heterostructure LaAlO$_3$/SrTiO$_3$ (LAO/STO) exhibits a wide range of physical phenomena that are mostly traced back to the rich properties of the STO system. When a thin layer (typically ≥4 unit cells) of LAO is grown on TiO$_2$-terminated STO, a two-dimensional electron gas spontaneously forms at the interface[1]. This system exhibits, under various conditions, superconductivity[2,3], magnetism[4], gate-tunable spin-orbit interactions[5,6], and tunable a metal-insulator transition[7]. The metal-insulator transition can be controlled by light[8,9], ion beam irradiation[10], applied back gate voltages[7], and c-AFM lithography[11,12].

Conductive AFM (c-AFM) lithography enables extreme nanoscale control of the metal-insulator transition in LAO/STO. The technique works by mediating a "water cycle"[13], selectively removing OH- from adsorbed water on the LAO surface, and allowing the remaining H$^+$ ions to modulation dope the LAO/STO interface. Applying a positive voltage to the c-AFM tip locally switches the LAO/STO interface to a conductive state, while negative voltages restore the insulating phase. The c-AFM technique has been widely used in exploring complex oxide systems, and many devices have been created at the 2DEG interface such as ballistic nanowires[14], electron waveguides[14], single-electron transistors[15], and novel forms of quantum matter [16,17].

c-AFM lithography, while capable of creating a wide range of nanoscale patterns, has practical limitations imposed by small scan ranges (~100 μm) and slow writing speed (~1 μm/s) of typical AFMs. Devices naturally decay in air on a time scale of hours[11,12], which also limits the complexity of devices that can be written.

To address these limitations and to enable more complex devices to be created, alternate approaches have been pursued, including parallel writing with AFM tip arrays[18]. Electron-beam lithographic (EBL) patterning enables larger devices with more complex layouts to be created. EBL is typically used with a resist such as PMMA[19], which is generally used with an additive or subtractive lithography step that is irreversible. High-energy electron beams are known to cause damage in highly insulating oxide materials, and the electron penetration for typical (>10 keV) electron acceleration energies can cause damage to the oxide material[20,21]. Moreover, the etching of material itself causes uncontrolled and, in many cases, undesirable behavior in the underlying material, which is sensitive to structural distortions.

In this Letter, we demonstrate an approach to reversible control of the metal-insulator transition in LAO/STO heterostructures, using an ultra-low voltage electron-beam lithography technique (ULV-EBL). This approach enables rapid, large-scale switching of the conductivity of the LAO/STO interface, with spatial precision comparable to c-AFM lithography, with no discernible changes to the topography of the LAO/STO structure. In addition to being significantly faster, the ULV-EBL technique enables patterning through the most widely investigated van der Waals (vdW) material, monolayer graphene, which points to an expansion of this approach to a wide range of 2D layered materials transferred to the LAO/STO platform.

## Experimental setup

A 3.4 unit-cells LAO is grown on top of TiO$_2$-terminated STO (001) substrate by pulsed laser deposition (PLD). Details of the growth technique are provide elsewhere[7]. The sample is initially insulating with a megaohm sheet resistance at the interface. Ti/Au (5nm Ti and 20nm Au) are deposited to the sample using standard photolithography to form "canvases" with low-resistance electrical contacts to the LAO/STO interface, as illustrated in Fig. 1a. The canvas is the area defined and surrounded by the electrodes within a $100 \times 100$ μm² clear area. The central canvas is designated for electron-beam exposure.

A commercial electron beam lithography instrument is used (Raith e-LiNE) with the electron acceleration reduced to 100 V. Unwanted electron-beam exposure to the canvas surface is avoided, and markers patterned on the edge of the sample are used for focus and alignment. The electron beam current is measured to be $I_e$ =195 pA and the write field is set to be 100 μm $\times$ 100 μm. The sample chamber vacuum is maintained at 1x10$^{-6}$ mbar during ULV-EBL writing and all electrical measurements, unless noted. Because LAO/STO samples are often light-sensitive, optical illumination inside the chamber, which is often used to adjust the sample stage position and e-beam gun position, is turned off after initial setup, and the sample is kept in dark during ULV-EBL writing. The conductance of the 2DEG devices is monitored *in situ* during the ULV-EBL writing process. After device patterning is complete, the sample is transferred to a dilution refrigerator, enabling electrical transport measurements to be performed over a temperature range of 50 mK to 300 K in magnetic fields up to 9 tesla. The c-AFM lithography is performed in an Asylum Research MFP-3D AFM in contact mode.

## Results

An experiment in which multiple lines are written between two interface electrodes by ULV-EBL and erased by the negatively biased AFM tip is shown in Figure 1. We define the real area dose as $D_r = (I_e \tau)/dsdl$ where $I_e$ is the beam current, $\tau$ is the dwell time, $ds = dl$ are the step size and line spacing. The dose used to expose the line is $D_r = 195$ μC/cm². The e-beam step size is $ds = dl = 10$ nm and line width $w = 50$ nm. As shows in Fig 1b, the nanowire is composed by dwell points of the electron beam. The writing speed is calculated to be 10 mm/s. Fig 1c shows the conductance jump from writing the first line. The on/off ratio is 153.7 and this ratio is dominated by the "off" value, which is more a function of the sample than the ULV-EBL. After writing, we transferred the sample into an AFM which took 1 hour 43 minutes to accomplish the transfer. During the transfer, the sample was kept in a gel box under ambient atmospheric conditions. The conductance was measured and subsequently "erased" using a c-AFM lithography. Figure 1d shows the final "cut" from the negatively-biased AFM tip in which the conducting channel is rendered insulating. The ability of c-AFM lithography to locally erase the nanostructure demonstrates that the writing process is reversible and suggests that the mechanism for controlling the conductivity is similar to c-AFM-based methods.

To understand the resolution of the writing process, nanowires with varying gaps of sizes $d$ are written using ULV-EBL (illustrated in Fig 2a) separated by $S = 5$ μm. By monitoring the conductance change with respect to the gap sizes, we observe significant changes in conductance, $\Delta G$, that appear when the gap size is in the range 5-20 nm (Fig 2c), indicating that the gap starts to be "covered" by the writing resolution at this range. Thus, we conclude that the resolution of the writing process is approximately 10 nm. As with c-AFM lithography, the actual resolution depends on the details of the writing parameters and possibly on variations between LAO/STO samples as well.

While room-temperature transport characteristics are themselves important, the ability of devices to remain conductive at low temperatures is critical for quantum device applications. Here we describe low-temperature transport properties of devices created by ULV-EBL. For these measurements, a four-terminal device is created (Fig. 3a) and transport measurements are performed in a dilution refrigerator (base temperature $T = 50$ mK). The conducting channel survives to $T = 50$ mK (Fig. 3b). The nanowire ($w = 2$ nm) shows evidence of superconducting behavior starting at $T_c = 200$ mK, consistent with other reports of superconductivity at the LAO/STO interface[2, 3] and in nanowire devices[22]. Fig. 3d shows an intensity map of differential conductance $dV/dI$ versus bias current $I$ and magnetic field $B$, showing values for critical current $I_c = 280$ nA extracting from Fig. 3c and upper critical field $H_c = 82$ mT. The asymmetry observed between positive and negative current is a known effect coming from the gating as semiconductors.

Previously, it has been shown that c-AFM lithography can control the LAO/STO metal-insulator transition through a monolayer of adsorbed graphene[23]. We tested whether it is possible to achieve the same type of control with graphene/LAO/STO heterostructures using ULV-EBL. Graphene grown by chemical vapor deposition is transferred onto LAO/STO using a method that is described elsewhere[24], and patterned into a rectangular shape. The graphene is electrically isolated from the LAO/STO interface electrodes, as illustrated in Fig. 4a and shown in the AFM scan in Fig. 4b. A close-up AFM scan (Fig 4c) shows that the graphene is conformal on the STO surface, with clear evidence of the 4 Å terraces of the STO underneath. A conductive channel with width $w = 1$ μm is exposed, causing a substantial conductance jump (Fig. 4d). Several tests were conducted consecutively to investigate the channel's conductance as a function of both the electron dose and nanowire width. For this experiment we choose a dose $D_0 = 975.75$ μC/cm$^2$. Also, we define $D$ as the dimensionless normalized dose factor. The real area dose ($D_r = D_0 D$) of the e-beam is then varied while writing a series of wires connected to the leads, the spacing of each experiment apart by 5 μm to minimize interactions. We observe that $\Delta G$ increases with the dose of the electron with the $w = 5$ nm wire (as shown in Fig 4e). The resulting conductance change $\Delta G$ versus the normalized dose factor $D$ for wires with different width is shown in Fig. 4f.

## Discussion

We now discuss possible mechanisms for reversible doping of the LAO/STO interface via ULV-EBL. We first consider mechanisms that are already known to result in doping of the STO layer. Oxygen vacancies, either in the STO or LAO layer [11], can shift the STO conduction band with respect to the Fermi energy. We find from CASINO Monte Carlo simulations (see supplementary material) that electrons are stopped before reaching the STO layer, most likely because of the low energy of the electron beam, and therefore direct e-beam-induced creation of oxygen vacancies in the STO can mostly be ruled out, although it is possible that they are created in the LAO layer. Electron-stimulated desorption of ions, specifically those which are believed to be desorbed using c-AFM lithography to selectively remove OH$^-$ species[13], are certainly candidate mechanisms[25]. The LAO surface is known to be covered by at least one monolayer of water which cannot be removed even under high vacuum conditions[26].

In summary, a new technique for reversible ULV-EBL-based patterning of the metal-insulator transition in LAO/STO and graphene/LAO/STO heterostructures has been demonstrated. The technique has sub-10-nm resolution and is capable of creating nanostructures that exhibit interesting behavior at low temperatures, including superconductivity. The fast writing speed and scalability of the ULV-EBL approach makes it well-suited to the development of much more complex families of quantum devices than before, including 2D simulation, arrays of THz and optical photodetectors, and graphene-based nanodevices.

# DATA AVAILABILITY STATEMENT

The data that support the findings of this study are available from the corresponding author upon reasonable request.

# ACKNOWLEDGEMENTS

The work at University of Pittsburgh was supported by ONR grant N00014-20-1-2481 and by Vannevar Bush Faculty Fellowship ONR grant N00014-15-2847. The work at University of Wisconsin-Madison (synthesis characterization of epitaxial thin film heterostructures) was supported by the US Department of Energy (DOE), Office of Science, Office of Basic Energy Science (BES), under award number DE-FG02-06ER46327.

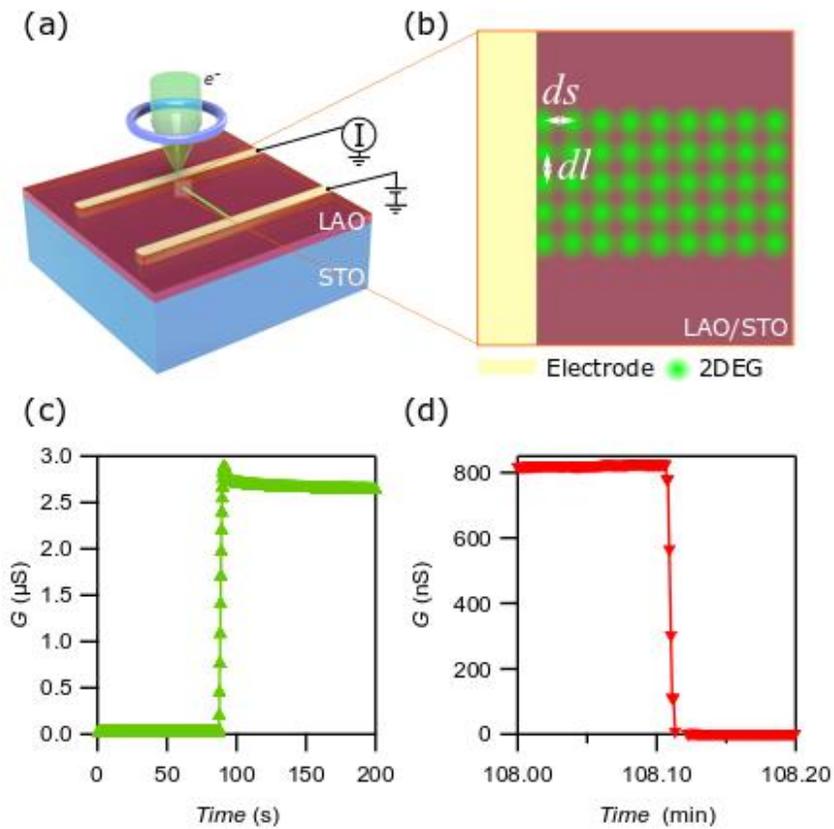

Fig1: **a**. Illustration of ULV-SEM on LAO/STO. **b**. Illustration of 2DEG nanowire geometry. The wire is composed of dwell dots of the electron beam. **c**. Conductance coming from connecting two interfacial electrodes by writing a nanowire with a 100 V e-beam. **d**. AFM erases the ULV-EBL created nanowire with a negatively-biased AFM tip. The sample is transferred to an AFM chamber after EBL writing. The transfer took 1 hour 43 minutes in ambient condition. Conductance drop results from cutting the existing nanowire.

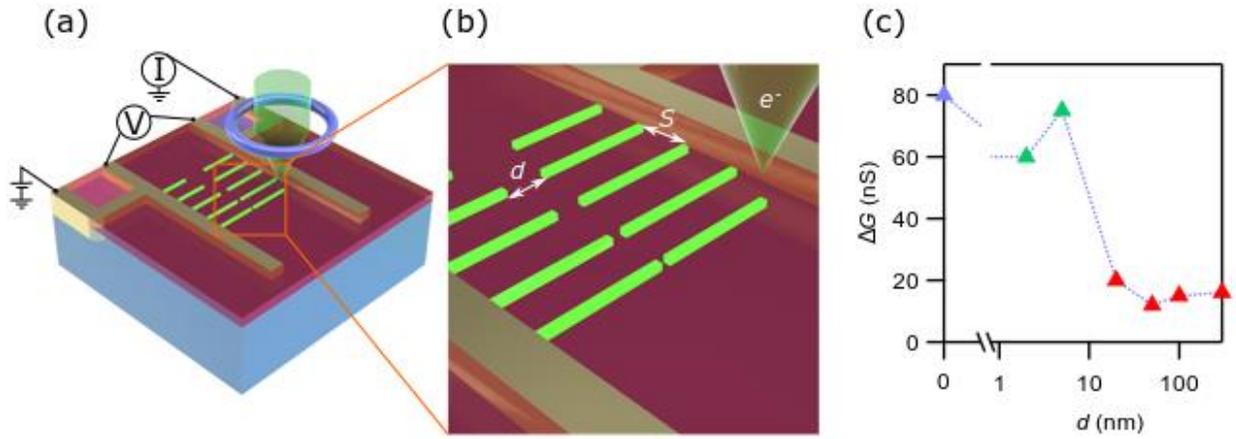

Fig2: **a.** Schematic diagram of writing nanowires with varying gaps of width $d$ in the middle. **b.** close-up illustrating the array of nanowires, gap separation $d$, and wire separation $S$. **c.** Measured conductance changes $\Delta G$ versus gap size $d$. A transition occurs between $d = 5$ nm and $d = 20$ nm. The red points label the insulating phase and green points are the conducting phase. The purple point ($d = 0$) represents an experiment where there is no gap.

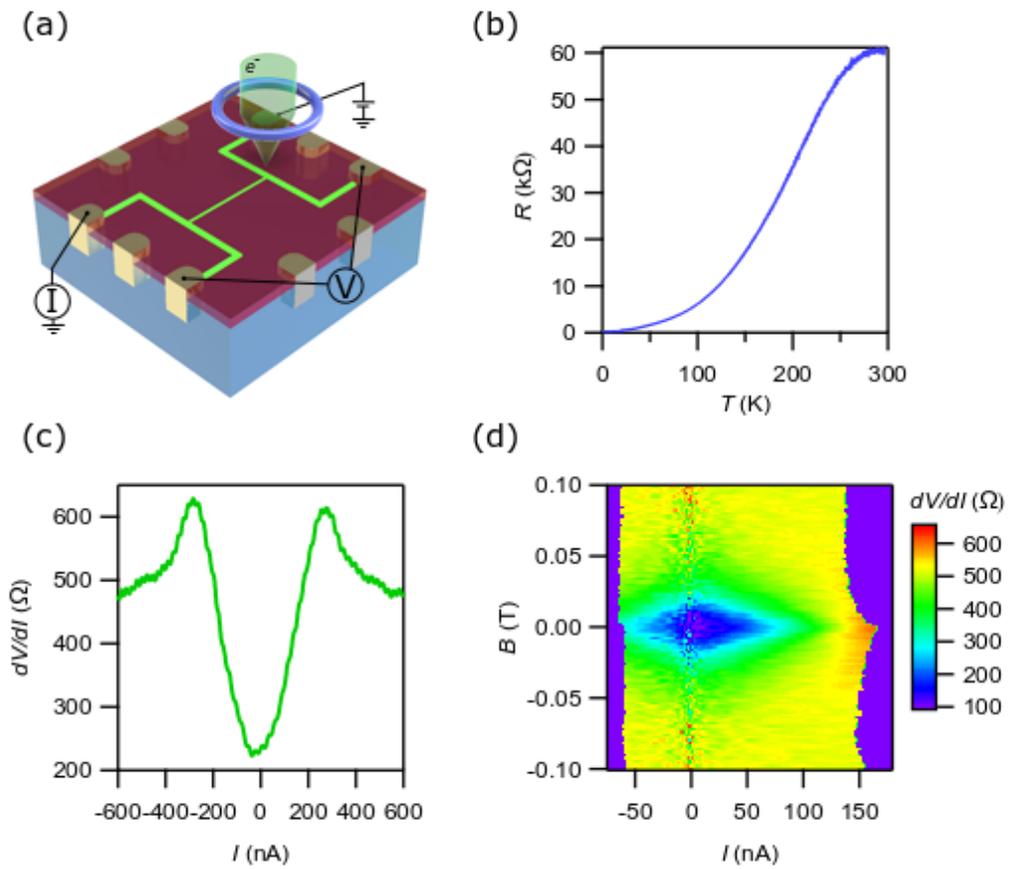

Fig3: **a**. Schematic diagram of the device. **b**. Cooldown curve of the $R$ with respect to the $T$. **c.** IV measurement at $T = 50$ mK showing a superconducting phase at $B = 0$ T. **d**. $dV/dI$ plotted as a function of $B$ and $I$. $T < 60$ mK

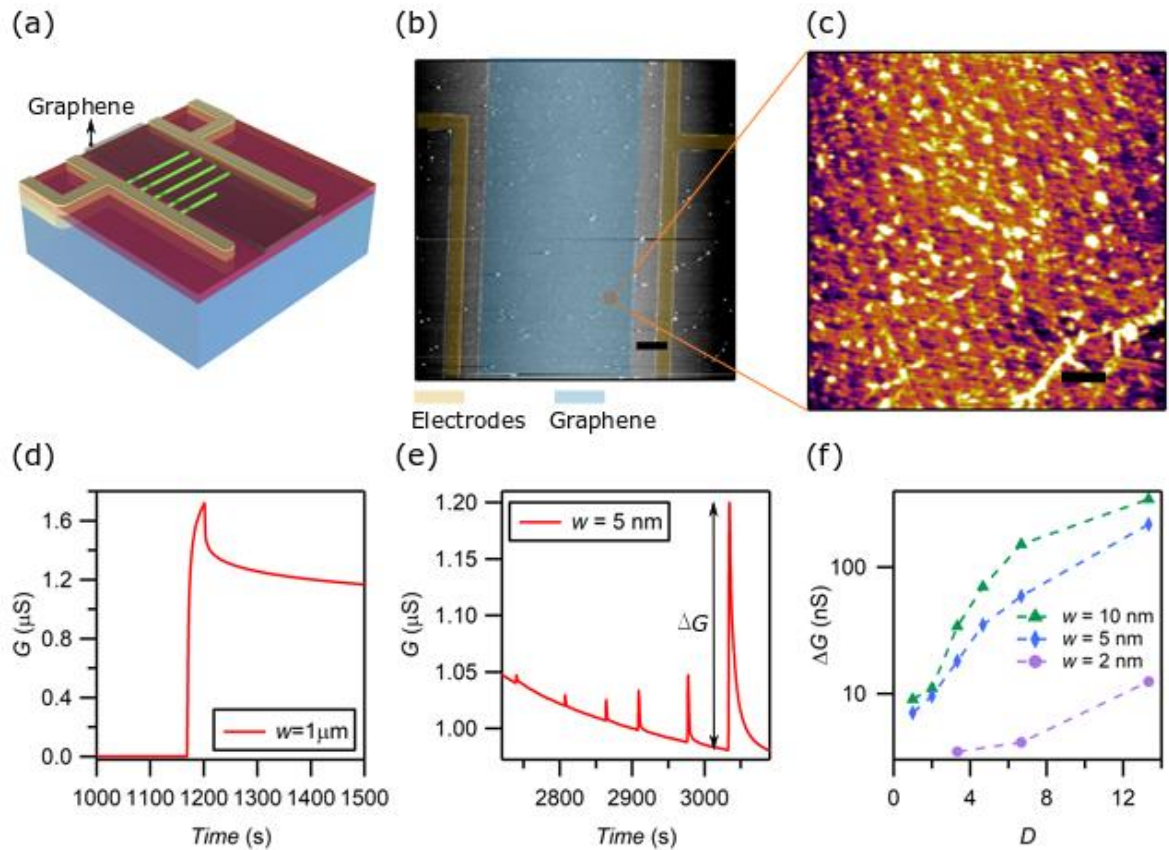

Fig4: **a**. Schematic diagram of device on graphene/LAO/STO. **b**. AFM profile of graphene on LAO/STO. The scale bar at the bottom right of the image corresponds to 10 μm. **c**. AFM image of graphene on LAO/STO. Terraces from the STO substrate are visible through the graphene. The scale bar at the bottom right of the scanning image corresponds to 500 nm. **d**. Conductance change when writing a 1-μm-wide strip connecting two interface electrodes. **e**. Conductance change while writing a series of 5-nm-wide wires with increasing dose factors. **f**. Conductance change $\Delta G$ with respect to normalized dimensionless dose factor $D$ for different width of wires.

**Supplementary material**

1. Monte Carlo simulation

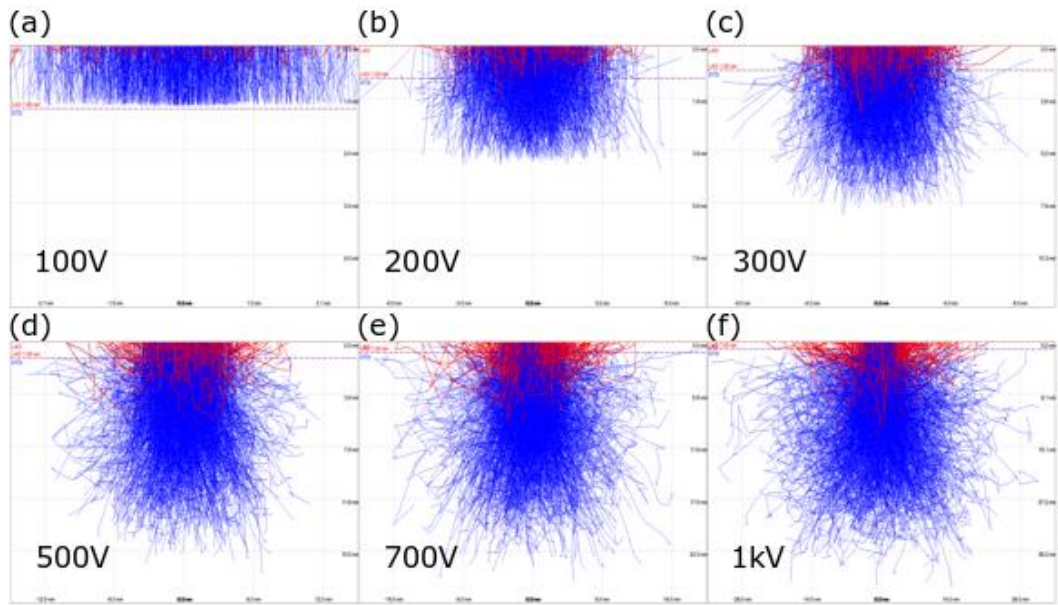

Figure S1. Monte Carlo simulation of electron trajectories in LAO/STO done by CASINO. Red lines are the backscattered electrons. Blue lines are the penetrated electrons. The red dashed line shows the interface between LAO and STO. The LAO is set to be 1.2 nm thick and the beam spot size is set as 5 nm. The accelerating voltage: (a) 100 V, (b) 200 V, (c) 300 V, (d) 500 V, (e) 700 V, (f) 1 kV.